\documentclass[reprint,letterpaper,twocolumn,floats,aps,prl,preprintnumbers,amsmath,amssymb]{revtex4}
\usepackage[T1]{fontenc}
\usepackage[latin9]{inputenc}
\usepackage{verbatim}
\usepackage{graphicx}

\begin{document}
\title{Patterns without patches: Hierarchical assembly of complex structures from simple building blocks}

\author{Michael Gr{\"u}nwald}
\affiliation{Computational Physics, University of Vienna, Sensengasse 8, 1090 Vienna, Austria}

\author{Phillip L. Geissler}
\affiliation{Department of Chemistry, University of California, Berkeley, California 94720}

\begin{abstract}
Nanoparticles with "sticky patches" have long been proposed as
building blocks for the self-assembly of complex structures.  The
synthetic realizability of such patchy particles, however, greatly
lags behind predictions of patterns they could form.  Using computer
simulations, we show that structures of the same genre can be obtained
from a solution of simple isotropic spheres, provided control only
over their sizes and a small number of binding affinities.  In a first
step, finite clusters of well-defined structure and composition emerge
from natural dynamics with high yield.  In effect a kind of patchy
particle, these clusters can further assemble into a variety of
complex superstructures, including filamentous networks, ordered
sheets, and highly porous crystals.
\end{abstract}

\maketitle

Living systems create and maintain their functional microscopic
organization through self-assembly, the spontaneous arrangement of an
initially unordered collection of biomolecular building blocks.
Mimicking this behavior in the laboratory with synthetic components
has proven to be a formidable challenge.  A close look at the agents
of self-assembly in living systems reveals a key aspect of the
problem: Most biomolecular objects interact through directionally
specific forces, so-called ``patchy'' interactions. Indeed, computer
simulations of model nanoparticles with attractive patches have
recapitulated much of the richness of nature's self-assembled
structures \cite{Zhang2005a,Vissers2013,Bianchi2011,Hagan2013,Ouldridge2011}.
Synthetic nanoparticles with controlled patchiness, however, are
largely unavailable in the laboratory, although
impressive progress has been made in specific cases
\cite{Wang2012,Chen2011,Pawar2010}.

In this letter, we consider a pragmatic question, though from a
theoretical perspective: Using only nanoparticle synthesis and
functionalization techniques that are standard today, can
self-assembled patterns be realized that share the complexity achieved
by biology's (and simulators') patchy components?  In particular, we
devise and demonstrate numerically a hierarchical strategy for this
purpose, which assumes control only over a few energies of interaction
between spherical particles, as well as their size. Such control
should be feasible in practice given well-established procedures to
decorate the exterior of nanoparticles with double-stranded DNA.

\begin{figure}
\includegraphics[width=0.42\textwidth]{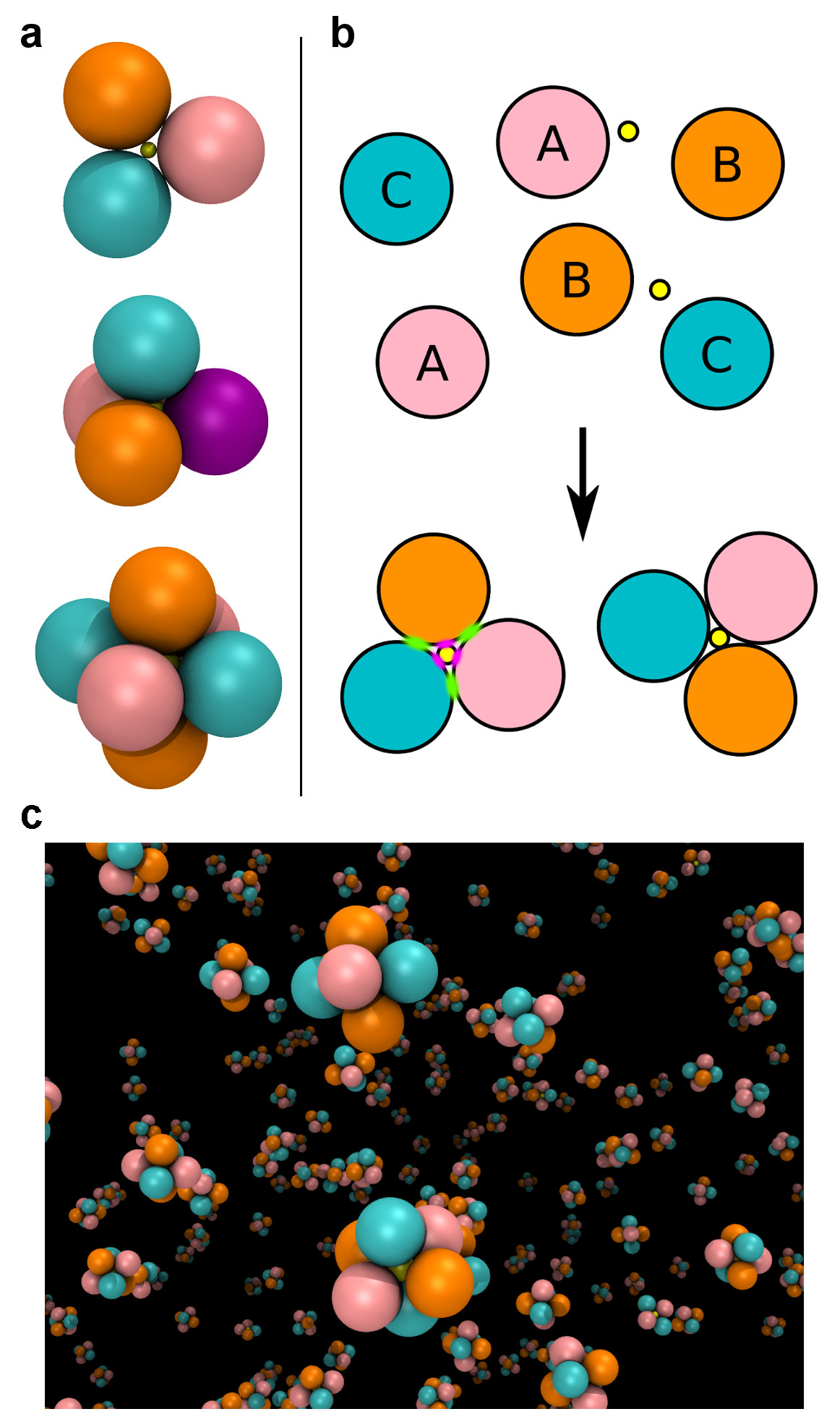}
\caption{\label{cluster_fig} {\bf Self-assembly of metaparticles.}
{\bf (a)} Three example metaparticles 
that can be prepared with high yield through appropriate choices of
attraction strengths and glue particle size. 
{\bf (b)} Strategy for controlling cluster size and composition. Purple contact points indicate strong attraction between glue particles (yellow)
and monomers [pink ($A$), orange ($B$), and blue ($C$)], which
dictates metaparticle size and shape.  Composition and connectivity
within a metaparticle are controlled through weaker interactions (green contact points) between monomers. In this example, all unlike particle
types attract one another; like types do not.
{\bf(c)} Snapshot from a simulation of $C_6$ metaparticle formation
(with $\bar{\epsilon}=3\,k_BT$ and $\epsilon_{\rm glue}=10\,k_BT$),
after equilibrium has been established. For clarity we show only
monomers that are bound to glue particles.  The yield of correctly
composed octahedral metaparticles in this case is $\approx90\%$.}
\end{figure}

Our scheme begins with a dilute solution of spherical particles, of
several types, that interact isotropically and over short
distances. With appropriate choice of the sizes and binding affinities
of these particles, we show that a nearly uniform population of
``metaparticles'' can emerge -- tightly bound clusters, comprising a
handful of spherical monomers, with defined composition and internal
structure, as illustrated in Figure \ref{cluster_fig}a. These objects
constitute a kind of patchy
nanoparticle, with nontrivial shape and an anisotropic arrangement of
monomers that can subsequently serve as sites for effectively
directional interaction. In the second stage of our scheme, the
emergent patchiness of metaparticles is exploited to spontaneously
generate large-scale superstructures, some of which are highly ordered
and reminiscent of biological assemblies.

The types of spherical monomers we have in mind are distinguished 
from one another
by
the strength of their interactions with other monomers. Specifically,
the potential energy of two monomers, of types $A$ and $B$, separated
by a distance $r$ is $u_{\rm rep}(r) + \epsilon_{AB}u_{\rm
  att}(r)$. The steric repulsion $u_{\rm rep}$
enforces volume exclusion, strongly penalizing separations below a
threshold value, $r \leq \sigma$. At distances near contact, the attractive potential $u_{\rm att}$
provides a favorable energy, $u_{\rm att}(r)\approx -1$ for $r \leq
\sigma + w$, and attenuates rapidly for $r > \sigma + w$.

The specific forms of 
these potentials are not important, only that $u_{\rm rep}$ sets a
well-defined particle diameter $\sigma$ and that $u_{\rm att}$ acts over a
short range $w\ll \sigma$ \footnote{We consider only attractive
  interactions with $\epsilon_{AB} >=0$. Negative values of $\epsilon$
  effectively increase the size of monomers. Because of the short
  range of interactions, this modification is subtle and has little
  effect on assembly dynamics.}. Colloidal nanoparticles with
surface-grafted DNA molecules provide one experimental realization of
this system, in which the complementary sequences of DNA strands
attached to monomers $A$ and $B$ encode the strength $\epsilon_{AB}$
of their attraction \cite{Knorowski2011,Geerts2010,DiMichele2013}.

Building finite-sized metaparticles from a macroscopic collection of
such monomers is not a trivial matter \cite{Manoharan2003,Licata2006a,Licata2008}. If, say, attractions among
monomers $A$, $B$, $C$, and $D$ provide the cohesive energy
maintaining the integrity of an $ABCD$ cluster, then additional
monomers of these four types will tend to bind at the cluster's
surface. Lacking constraints on monomer valency, it is not clear how
to design against unbounded growth of a close-packed crystal. 

Indeed, an extensive search
through possible combinations of binding affinities 
did not yield self-limiting growth
of small clusters in computer simulations.

To prepare metaparticles we instead adapted an approach devised to
build finite clusters of identical particles \cite{Soto2002,Fan2011}. Here,
cluster size and geometry are dictated by introducing an additional
kind of particle with smaller diameter $\sigma_{\rm glue} <
\sigma$. This ``glue particle'' attracts all other monomers strongly
(with contact energy $\epsilon_{\rm glue}$) and over short range
($w_{\rm glue}$).  For appropriate size combinations, the propensity
to maximally coordinate each glue particle determines with great
precision the structure of its shell of monomers, as illustrated in
Figure \ref{cluster_fig}b. 
Of the many convex polyhedra that can be obtained in this way (denoted
$C_n$, where $n$ is the number of shell particles), we
focus exclusively on triangles ($C_3$), tetrahedra ($C_4$), and
octahedra ($C_6$) illustrated in Figure \ref{cluster_fig}a. Unlike larger shapes, we can control the
arrangement of monomer types within these shells through the set of
attraction strengths ${\boldsymbol{\epsilon}} \equiv
\{\epsilon_{AA},\epsilon_{AB}, \ldots\}$, as discussed below.

The only threat to self-limiting growth in this scenario is the
possibility that two glue particles bind to overlapping sets of shell
monomers. This errant growth can be made irrelevant by working at low
concentration of glue particles. Alternatively, they could be endowed
with longer-range repulsions. Our simulations follow the latter
approach, with glue particles repelling one another through a screened
Coulomb interaction (see Methods).

\begin{figure*}
\includegraphics[width=0.95\textwidth]{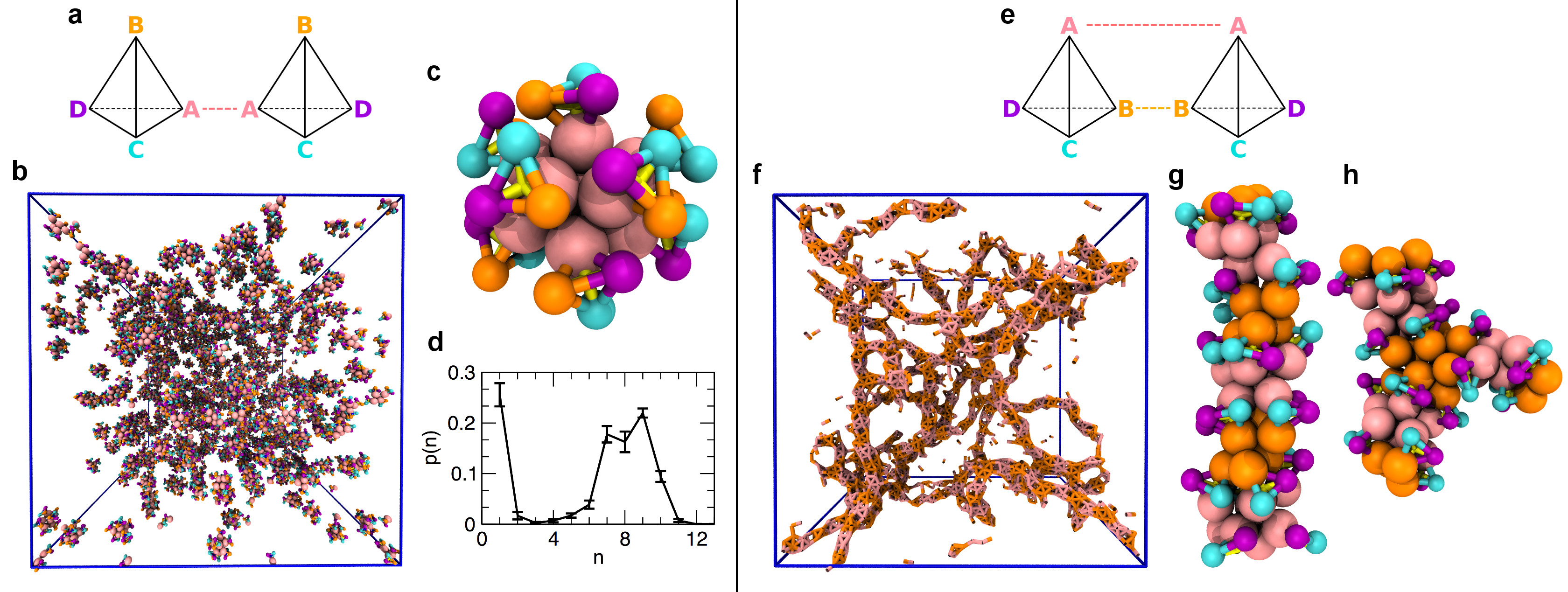}
\caption{\label{micfil_fig} 
{\bf Zero- and one-dimensional assemblies} 
(Left) Micelle-like
superstructures.  Attractions of type $[AA]$ (indicated by the dashed
pink line in {\bf (a)}) between $C_4$ metaparticles yield
self-limiting growth of superclusters whose interiors are dense in $A$
monomers.  {\bf (b)} Snapshot from a molecular dynamics simulation,
showing a dispersed collection of micelles.
{\bf (c)} 
An example micelle comprising 9 $C_4$ clusters.  {\bf (d)} Normalized
histogram of the number of metaparticles within each micelle. 
(Right) Filamentous assemblies.  Attractions of type $[AA,BB]$
(indicated by the dashed pink and orange lines in {\bf (e)}) between
$C_4$ metaparticles yield percolating networks of branched filaments.
{\bf (f)} Snapshot from simulation, showing only $A$--$A$ and $B$--$B$
bonds.  An example filament segment {\bf (g)} and branch point {\bf
  (h)}. The filament core consists of alternating $A$-rich and
$B$-rich regions; non-attracting monomers of type $C$ and $D$ form a
loose shell around the core.  (In panels {\bf (b)}, {\bf (c)}, {\bf (g)}, and {\bf (h)}  
non-attracting monomers are shown in smaller size for clarity.)  }
\end{figure*}

\begin{figure*}
\includegraphics[width=0.95\textwidth]{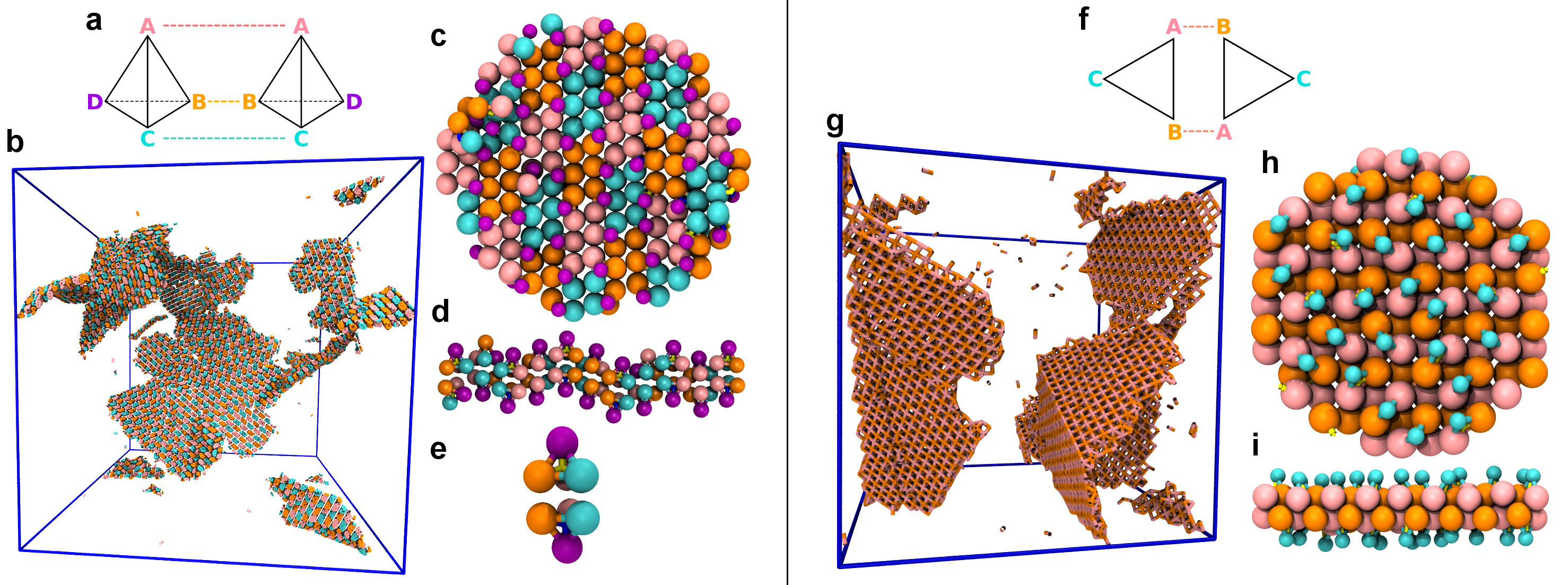}
\caption{\label{sheets_fig} {\bf Two-dimensional assemblies} (Left)
  Bilayer sheets of $C_4$ metaparticles.  Attractions of type
  $[AA,BB,CC]$ (indicated by the dashed pink, orange, and blue lines
  in {\bf (a)}) yield sheets two metaparticles thick.  {\bf
    (b)} Snapshot from a molecular dynamics simulation, showing a
  collection of sheets linked at grain boundaries. (Only $A$--$A$, $B$--$B$, and $C$--$C$
bonds are shown.) {\bf (c)} Top view
  of an example sheet, whose attracting monomers are arranged in
  groups of 12 in a regular pattern. Metaparticles with different
  chirality alternate in the sheet plane. {\bf (d)} Side view of the
  same example sheet, highlighting its corrugation. 
(Centers of enantiomers are shown yellow
and blue.)  
{\bf (e)} Each cluster is bound to its mirror image in the
opposite layer of the sheet. 
(Right) Bilayer sheets of $C_3$ metaparticles.  Attractions of type
$[AB]$ (indicated by the dashed lines in {\bf (f)}) yield
sheets that are flat and highly ordered.  {\bf (g)} Snapshot from a
molecular dynamics simulation, showing only the bonds between $A$ and
$B$ monomers.  Of the two disconnected sheets, one
exhibits extended kink defects.  Top {\bf (h)} and side views {\bf (i)}
of an example sheet highlight the face-centered-cubic-like bonding
geometry. (In panels {\bf (c)}--{\bf (e)}, {\bf (h)}, and {\bf (i)} monomers are
shown in smaller size for clarity.)}
\end{figure*}

\begin{figure*}
\includegraphics[width=0.95\textwidth]{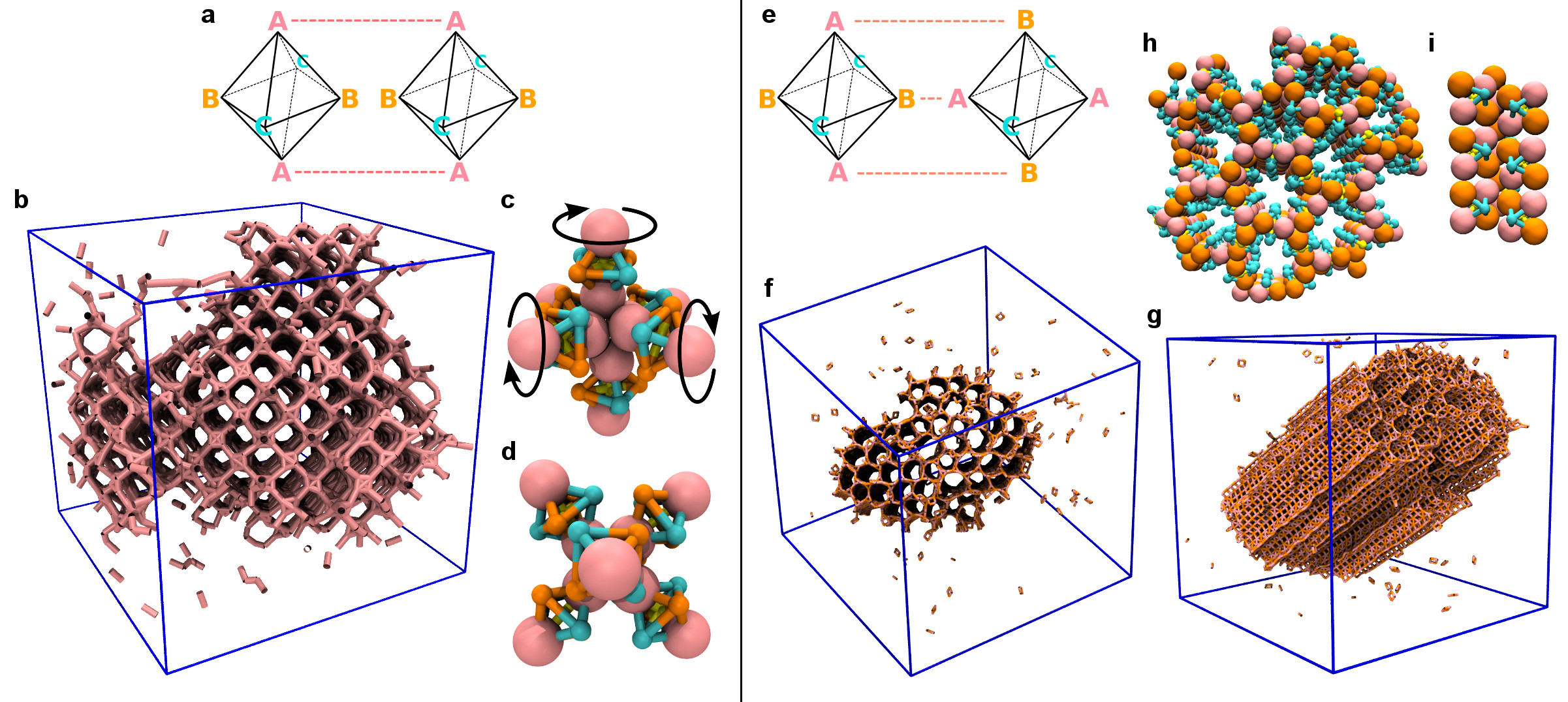}
\caption{\label{crystals_fig} 
{\bf Three-dimensional assemblies.}
(Left) A rotator phase with cubic symmetry.  Attractions of type
$[AA]$ (indicated by the dashed pink lines in {\bf (a)}) between $C_6$
metaparticles yield a cage-like supercrystal. {\bf (b)} Snapshot from
a molecular dynamics simulation, showing only the bonds between $A$
monomers.  ({\bf (c)} and {\bf (d)}) An example vertex of the
superlattice viewed from two different perspectives.  Each vertex
involves six $C_6$ clusters, bound through $A$-$A$ attractions in a
cross-like geometry. Metaparticle centers occupy the Wyckoff $3d$
positions of a cubic crystal with space group 221 (Pm$\bar3$m). The
maximal packing fraction of the crystal is $\approx 0.42$. Each
metaparticle can rotate freely around its $A$-$A$ axis, as
indicated by arrows.  (Right) Honeycomb supercrystal.  Attractions of
type $[AB]$ (indicated by the dashed pink lines in {\bf (e)}) between
$C_6$ metaparticles yield a superstructure featuring extended channels
arranged in a hexagonal pattern.  ({\bf (f)} and {\bf (g)}) Simulation
snapshot, showing only the bonds between $A$ and $B$ monomers, viewed
from two different perspectives.  ({\bf (h)} and {\bf (i)}) Excerpts
from the honeycomb viewed along the channel axis and from its side.  Channels are defined by
circular arrangements of six $C_6$ clusters; non-attracting particles
of type $C$ point towards the center of the pore.  (In panels {\bf (c)}, {\bf (d)}, {\bf (h)}, and {\bf (i)} non-attracting monomers are
shown in smaller size for clarity.)}
\end{figure*}

The crux of making well-defined metaparticles 
lies in dictating the identities of monomers at each vertex of the
shell. Out of the many
possible shell compositions, illustrated for $C_4$ clusters in Supplementary Figure 1, 
one must be represented with dominant statistical weight. 
As one challenge to this task, $\epsilon_{\rm glue}$ must be
sufficiently weak that binding is reversible, so that inevitable
mistakes in cluster composition can be corrected in reasonable time;
cluster integrity may be compromised as a result.  More subtly,
attractions among shell particles must be sufficiently weak that they
do not macroscopically condense. This constraint limits the extent to
which one shell composition can dominate energetically over others. It
is not obvious that these competing requirements can all be satisfied
with a single choice of $\epsilon_{\rm glue}$ and
$\boldsymbol{\epsilon}$.

Approximate analytical calculations, as well as explicit Brownian
dynamics simulations, indicate that high yields of certain
metaparticles can in fact be made in this simple fashion (see Supplementary Methods and Supplementary Figure 2).
As a straightforward design, we set $\epsilon_{ij}=\bar{\epsilon}$ 
if monomer types $i$ and $j$ 
make contact in a desired cluster, and $\epsilon_{ij}=0$
otherwise \cite{Hormoz2011}. Due to the limitations of short-ranged, pairwise, and
isotropic interactions, this scheme does not permit access to all
metaparticle compositions. For example, we cannot generate a pure
population of octahedra with more than three monomer types. 
However, the metaparticles shown in Figure \ref{cluster_fig}a can be
prepared with high fidelity through spontaneous dynamics of initially
dispersed monomers and glue particles. (See Figure \ref{cluster_fig}b;
simulation details are given in the Methods section.)  In particular,
we achieve maximum yields of 94\%, 78\%, and 98\% for $C_3$, $C_4$,
and $C_6$ clusters respectively. In the assembly of $C_4$
clusters, the two enantiomers of an $ABCD$ tetrahedron must appear
with macroscopically identical concentrations -- in simulations of
superstructure assembly described below we consider the racemic
mixture. In all cases, the emergent patchiness of clusters is
sufficient to generate a rich variety of self-assembled
superstructures.

To induce further assembly among many metaparticles, it is necessary
at this point to modify the strengths of attraction
$\boldsymbol{\epsilon}$ between their constituent monomers, which now
act as sticky patches for interactions between distinct clusters. To
avoid consequent changes in metaparticles' internal structure, it is
further necessary to render the glue particle bonds irreversible. Both
of these tasks have been accomplished in other contexts using 
techniques of DNA nanotechnology \cite{DiMichele2013}. Fortunately, elaborate combinations
of monomer attraction are not needed at this stage to assemble complex
patterns. On the contrary, introducing substantial attractions between
more than one or two monomer types typically allows only close-packed
crystals or amorphous solids as products of 
assembly. We have instead obtained interesting and varied assembly
when $\epsilon_{ij}=0$ for all monomer-monomer interactions except:
(i) self-attraction of one monomer type, i.e., $\epsilon_{AA}> 0$ (a
design we denote $[AA]$); or (ii) self-attraction of two types,
$\epsilon_{AA}> 0$ and $\epsilon_{BB}> 0$ (denoted $[AA, BB]$); or
(iii) a single cross-interaction, $\epsilon_{AB}> 0$ (denoted
$[AB]$). In one special case the design $[AA, BB, CC]$ was also
conducive to nontrivial pattern formation. As is generally the case
with patchy nanoparticles, the dynamical fate of assembly is very
sensitive to the magnitudes of these attractions
\cite{Hagan2006,Jack2007,Grant2011}.  
We explored a range of values of $\epsilon_{AA}$, $\epsilon_{BB}$, etc.
for each structure and report here on choices that yielded the most
reproducible and defect-free assemblies. For an attraction range
$w \approx 0.05 \sigma$, well depths of a few
$k_BT$ appear to be optimal in all cases (see Methods). 
For some designs
the energetic range between impractically slow growth and
extensively defective aggregation 
is
as narrow as $0.2\,k_{\rm B}T$. 
The superstructures described below
were assembled from metaparticles of uniform composition; 
their quality was 
only slightly degraded by including defective metaparticles
at the levels
indicated by simulations of cluster formation.

Given that metaparticle structures are highly symmetric, and that the
size of effective patches is prescribed by the monomer diameter, one
might expect the variety of patterns that can be assembled from 
%our
the
$C_3$, $C_4$, and $C_6$ clusters 
of Figure \ref{cluster_fig}a
to be meager and easily
anticipated. These objects, however, are more complex building blocks
than spheres decorated with symmetric interaction sites. Indeed,
particle shape can be a critical factor in self-assembly, strongly
influencing the structure of thermodynamic ground states as well as
their kinetic accessibility\cite{Henzie2012,Miszta2011,Damasceno2012}. 

An interplay between packing and directional attraction is important
in even the simplest assembly we observed in Brownian dynamics
simulations of interacting metaparticles. (Simulation details are given in the Methods section.)
With only one mode of
self-attraction ([$AA$]) among monomers in different $C_4$ clusters,
the attracting species $A$ tends to aggregate to the extent allowed by
volume exclusion due to the rest of the cluster.  Zero-dimensional,
micelle-like superstructures thus naturally emerge (see Figure \ref{micfil_fig}a), and
like conventional micelles they are not uniform in size. Because of metaparticles' anisotropic shape,
the average size and polydispersity of these superstructures are the
result of a complicated competition between the energetic drive to
expand the $A$-rich cores and the entropic cost of packing (bumpy)
tetrahedra at locally high density. 

Adding a second self-attraction to $C_4$ species ([$AA, BB$])
effectively encourages aggregation within and among the micellar
superstructures just described. Due to constraints of packing and
stoichiometry, $B$ monomers on the micelle exterior segregate to
opposite poles, where they can bind to the poles of other micelles.
This linear motif may extend indefinitely, generating one-dimensional
super-filaments with an internal pattern of alternating $A$-rich and
$B$-rich cores, as shown in Figure \ref{micfil_fig}b. The fluctuations responsible for
micelle size variation in the $[AA]$ case here produce local defects
in core thickness and exposure. Some of these defects cause sufficient
exposure of the cylindrical core to allow filament branching.  At high
metaparticle density, percolating networks of filaments reminiscent of
biopolymer gels result \cite{Fletcher2010}. 

Adding a third attraction to this scenario ([$AA, BB, CC$]) once again
increases the dimensionality of assembled superstructures. Only
monomers of type $D$ are inert in this case, and they can be
sequestered to the opposing faces of an ordered sheet that is two
metaparticles thick, as illustrated in Figure \ref{sheets_fig}a. In its ideal form this superstructure features
macroscopic lines of isochiral clusters, alternating with lines
of their enantiomers. Such chiral micropatterning might provide a
basis for engineering unusual optical properties. Sheets forming on the time scale of our simulations possess
a significant number of defects that define grain boundaries between domains of different orientation.

Triangular metaparticles were similarly observed to form sheets (with $[AB]$ and $[AA,BB]$ attractions) and
micelles (with $[AA]$ attractions), as illustrated in Figure \ref{sheets_fig}b and Supplementary Figure 3.
Octahedral clusters, on the other hand, generate exotic three-dimensional
superlattices. In one such crystal ($[AA]$), $C_6$ metaparticles
maintain complete rotational freedom about the axis connecting
attractive monomers (see Figure \ref{crystals_fig}a).  Another crystal (obtained with $[AB]$ and $[AA,BB]$) is highly porous, with a
packing fraction of $\approx 0.12$, and is traversed by hexagonal
channels (see Figure \ref{crystals_fig}b and Supplementary Figure 4). Both offer intriguing design possibilities for molecular adsorption and metamaterials.

While this survey of assemblies is not necessarily exhaustive, we
believe it to be thorough for metaparticles accessible with high yield
through the procedures we have described. Mixtures of metaparticles
with varying composition are 
in fact easier to prepare, at least in the proportions dictated by
their thermodynamic stabilities. 
Such mixtures 
expand the range of superstructures that can be achieved through
our hierarchical protocol.
As one example,
$C_6$ clusters of uniform
composition $ABCBCD$ 
cannot be prepared with high
yield from our strategy. However, a mixture of 
$ABCBCD$, $ABCBCA$, and $DBCBCD$ clusters
(in average proportions 2:1:1) can be straightforwardly generated,
specifically
by adding a fourth monomer type $D$
to the collection of monomers that would otherwise form
pure $ABCBCA$ clusters.
In simulations, 
this mixture
assembles into yet another distinct porous
supercrystal, 
shown
in Supplementary Figure 5. 
Many more such
scenarios are 
possible and, more importantly, should be straightforward
to realize using existing synthetic technologies.

\subsection{Methods}

\subsubsection{Pair potential}

The repulsive and attractive interaction potentials
used in our simulations have the specific forms
$$
u_{\rm rep}(r) =
 \left\{ \begin{array}{cl} 
k_{\rm B}T [1 + 4(\tilde{r}^{-12}- \tilde{r}^{-6}) ], & r < \sigma \\
                        0, & r \geq \sigma\,,
		\end{array}
	\right. 
$$
where
$$
\tilde{r}(r) = \frac{r-\sigma}{\alpha_{\rm rep}}+2^{1/6}\,,
$$
and
\begin{multline*}
u_{\rm att}(r) =\\
 \left\{ \begin{array}{cl} 
\frac{1}{2}\left[\tanh\left(\frac{r-(\sigma+w)}{\alpha_{\rm att}}\right) - \tanh\left(\frac{w}{\alpha_{\rm att}}\right)\right], &
r < \sigma+2w \\
                        0, & r \geq \sigma + 2w\,.
		\end{array}
	\right.
\end{multline*}
The repulsive part, a shifted Lennard-Jones potential whose steepness
is set by the length scale $\alpha_{\rm rep}$, vanishes continuously at $r=\sigma$.
The attractive part, a well of approximately unit depth whose
steepness near $r=\sigma+w$ is set by the length scale $\alpha_{\rm att}$, vanishes
continuously at $r=\sigma+2w$. Examples of the total interaction
potential are plotted in Supplementary Figure 6.  The standard mixing
rule $\sigma = (\sigma_1 + \sigma_2)/2$ was used to determine
interactions between particles with different diameters $\sigma_1$ and
$\sigma_2$.

\subsubsection{Simulations}
In all simulations we adopt $k_{\rm B}T$ as a unit of energy, monomer
diameter $\sigma$ as a unit of length, monomer mass $m$ as a unit of
mass, and $\tau = \sqrt{m \sigma^2/k_{\rm B}T}$ as a unit of time.
Our systems are then specified by the following dimensionless
parameters: (a) glue particle diameter, $\sigma_{\rm glue}/\sigma$,
which we set to $\sqrt{4/3}-1$, $\sqrt{3/2}-1$, and $\sqrt{2}-1$, for
$C_3$, $C_4$, and $C_6$ clusters, respectively; (b) glue particle
mass, $m_{\rm glue}/m = (\sigma_{\rm glue}/\sigma)^3$, proportional to
its volume; (c) monomer friction coefficient $\gamma \tau/m = 10$; and
(d) glue particle friction coefficient $\gamma_{\rm glue} \tau/m = 10
\sigma_{\rm glue}/\sigma$, proportional to its diameter.  Dynamics
were advanced by numerically integrating the underdamped Langevin
equation, as implemented in the HOOMD-blue simulation package
\cite{HOOMD}.

\subsubsection{Cluster assembly}
For simulations of metaparticle formation, we set $w=0.035\,\sigma$,
$\alpha_{\rm rep}=0.2\,\sigma$, and $\alpha_{\rm att}=0.01\,\sigma$.  Initial
conditions were constructed by randomly placing 500 glue particles and
1000 monomers (2000 in the case of $C_6$ clusters) of each type in a periodically replicated cubic
simulation box at packing fraction $0.005$.  Trajectories of length
$10^4\,\tau$ were generated with an integration time step $\Delta t =
10^{-4}\,\tau$.  Binding of monomers to multiple glue particles was
suppressed by a pairwise repulsion $u_{\rm glue-glue}(r)= 40 k_{\rm
  B}T \,e^{-r/\sigma}(r/\sigma)^{-1}$.  We calculate assembly yields as $N_\mathrm{t}/N$, where $N_\mathrm{t}$ is the number of clusters with desired composition and $N$ is the total number of clusters with the maximum number of monomers. Clusters that are not fully assembled are disregarded, as their population can be made negligible by an
appropriate choice of $\epsilon_{\rm glue}$. Maximum yields were achieved with
$\epsilon_{\rm glue} = 10 \,k_{\rm B}T$ and $\bar\epsilon = 4\,k_{\rm
  B}T$ for $C_4$ and $C_6$ metaparticles, and $\bar\epsilon =
4.4\,k_{\rm B}T$ for $C_3$ clusters.

\subsubsection{Superstructure assembly}
Simulations of the second stage of assembly included 1000 to 8000
metaparticles, initially placed on a simple cubic lattice at densities
between $0.04\,\sigma^{-3}$ and $0.01\,\sigma^{-3}$.  Metaparticles
were treated as rigid bodies \cite{HOOMD_rigid}.  Monomer interaction
parameters were set as $w=0.075\,\sigma$, $\alpha_{\rm
  rep}=0.3\,\sigma$, and $\alpha_{\rm att}=0.02\,\sigma$ (which allow use of a
larger integration time step $\Delta t = 0.005\,\tau$). Glue particle
repulsions were omitted at this stage. Time was advanced in each
assembly trajectory by $5\times10^5\,\tau$.

The following attraction strengths
resulted in the structures depicted in Figures \ref{micfil_fig},
\ref{sheets_fig}, and \ref{crystals_fig}:
\begin{itemize}
\item $C_4$ micelles: $\epsilon_{\rm AA}=5.8\,k_{\rm B}T$
\item $C_4$ filaments: $\epsilon_{\rm AA}=\epsilon_{\rm BB}=4.5\,k_{\rm B}T$
\item $C_4$ sheets: $\epsilon_{\rm AA}=\epsilon_{\rm BB}=\epsilon_{\rm CC}=3.7\,k_{\rm B}T$
\item $C_3$ sheets: $\epsilon_{\rm AB}=3.7\,k_{\rm B}T$
\item $C_6$ cubic rotator phase: $\epsilon_{\rm AA}=4.7\,k_{\rm B}T$
\item $C_6$ hexagonal channels: $\epsilon_{\rm AB}=4.15\,k_{\rm B}T$
\end{itemize}
These values resulted in structures of the highest quality in our
simulations. We note, however, that different values are likely to be
optimal for different choices of pair potential (in particular, for a
different range $w$ of attraction, as discussed in Supplementary
Methods), and for assembly trajectories that are substantially longer
than the time scales accessible with current hardware.  
 
Images of clusters and assemblies were rendered with VMD \cite{VMD}.

\subsection{Acknowledgments}
We thank Michael Brenner, Todd Gingrich, Sharon Glotzer, and Patrick Varilly for useful discussions. This work was supported by the Austrian Science Fund (FWF) under Grant J 3106-N16. Calculations were in part performed on the Vienna Scientific Cluster (VSC).  

%\bibliographystyle{mybibstyle}
%\bibliography{/Users/gruen/bibtex/library}

\end{document}